# Strip Searching for Supernovae

Ermanno F. Borra,

Département de Physique, Centre d'Optique, Photonique et Laser, Université Laval, Canada G1K 7P4

Email: borra@phy.ulaval.ca

LMT WEB PAGE: http://astrosun.phy.ulaval.ca/lmt/lmt-home.html

Subject headings: Supernovae, Cosmology

2## ABSTRACT

The survey of a strip of sky carried out with a zenith-observing 4-m liquid mirror can discover and observe several thousand supernovae/year to $z = 1$, making it quite competitive with other proposed surveys. Because LMs are inexpensive, and easy to build, a SNeIa survey can begin rapidly. As documented in the published literature, the present technology, and its cost, has been proven in both the laboratory and in observatory settings and is ready, at a very modest cost, for a 4-m class telescope. As a bonus, the coadded data would survey 150 square degrees of sky to magnitudes > 27, as deep as the deepest optical surveys but in a far larger area of sky. There presently is a window of opportunity to make fundamental discoveries by exploiting the cost advantage of this powerful novel technology.
## 1. INTRODUCTION

There currently is intense interest to observe Type Ia supernovae because they are bright standard candles. The reasons to do this are well-known (e.g. Huterer & Turner 2001 and references therein) so that there is no need to repeat them here. Several surveys have been proposed (e.g. Wang 2000 and references therein) to observe of the order of one thousand supernovae to redshifts as large as $z = 1.5$. Because they involve a considerable number of nights on expensive large (or space) telescopes these ambitious surveys will be expensive and will, in all likelihood, not be carried out in the next few years. Other proposals want to reserve large batches of time on existing 4-m telescopes.

This would have two major inconveniences: The number of nights one could realistically obtain are a small fraction of what is available on a dedicated 4-m telescope and would greatly disrupt other worthwhile scientific projects. The purpose of this article is to point out that the survey of a strip of sky carried out with an inexpensive zenith-observing 4-m LMT can start soon and will observe several thousand supernovae/year to z = 1.

## 2. PERFORMANCE OF THE TELESCOPE

A zenith telescope that tracks with a CCD in the TDI mode observes with an integration time given by the time it takes an object to drift across the CCD detector. The nightly single-pass time (in seconds) is given by

$$t = 1.37 \times 10^{-2} \, n \, w \, /(f \cos(lat)), \qquad (1)$$

where n is the number of pixels along the read-out direction of the CCD, w the pixel width (microns), f the focal length of the telescope (meters) and *lat* is the latitude of the observatory. TDI tracking with a zenith telescope aberrates the PSF but there are correctors that can remove most of it (Hickson & Richardson 1998).

A 4-m diameter f/2 LMT equipped with a 4KX4K CCD having 15-micron pixels ( 0.4 arcsec/pixel) obtains an integration time of 120-seconds/nightly pass. Table 1 gives the limiting magnitudes in B, R and I estimated with the IRAF task CCDTIME available on the NOAO web site, assuming the CCD prime focus camera at the Kitt Peak 4-m telescope. The zenith observations assume 1 arcsecond seeing and a 7-day-old moon.

A figure (e.g. figure 1 in Borra 1987) that converts equatorial into galactic coordinates shows the strip of sky sampled by a zenith telescope at a given site. The strip of sky follows a straight line of constant declination equal to the terrestrial latitude of the observatory. At a terrestrial latitude of + 30 degrees, it rises from the galactic plane, climbs through the North Galactic Pole and falls back to the plane; at -30 degrees it goes through the bulge of our galaxy and the SGP.

The width of the strip of sky observed by the CCD is given by

$$S = 0.206 \, n \, w / f, \quad (2)$$

where S is expressed in arcseconds, w is the pixel width in microns, and n is the number of pixels in the direction perpendicular to the scan. With a 4KX4K CCD mosaic having 0.4 arcseconds pixels, the strip of sky is 26 arcminutes wide. Using spherical trigonometry on the celestial sphere, one finds that it observes 146 square degrees, 72 square degrees of which are "extragalactic" regions of low galactic obscuration having galactic latitudes > 30 degrees.

**3. A SUPERNOVA STRIP SURVEY**

A zenith telescope observes the same region of sky night after night and measures the fluxes of all objects in it every clear night. Faint objects can be observed in the visible region of the spectrum for two weeks of moonless nights a month; there is no restriction for observations redder than R. The sampling time is well matched to the time scales of SN light curves.

Table 2 gives the apparent magnitudes at maximum light of Type Ia supernovae as a function of redshift (Schmidt et al. 1998). Comparison of Tables 1 and 2 shows that a single nightly pass can detect a SNIa near maximum light to z=0.8.

Furthermore, considering the 1+z time dilation factor, one can bin 6 nights at z>0.5, going 1 magnitude deeper, and to z > 1. Binning 6 nights in I is quite realistic if one considers that the I limiting magnitudes are essentially independent of the phase of the moon. Light decay time corrections to canonical peak luminosities need observations to 15 days in the rest frame, at which time the luminosity has decreased about 2 magnitudes below the peak, allowing one to observe useful light curves to z = 1.0.

SN rates are uncertain. Pain et al. (1996) estimate the rate of Type Ia supernovae at z ~ 0.4. They predict 34.4 (+23.9, -16.2) events/year/square degree for magnitudes in the range 21.3 < R < 22.3, corresponding to 0.3<z<0.5. Section 2 shows that a survey with a 4KX4K CCD covers an extragalactic strip of sky ( bII > 30 degrees) having a surface of 72 square degrees so that there should be about 3000 events per year in the strip. Approximately 2/3 of the objects would be unobservable in a given night because they would lie outside of the nightly strip monitored, reducing the number to about 1000 SNe Ia per year having 0.3<z<0.5. Table 3 estimates the number of SNe Ia in other redshift ranges by naively extrapolating with the differential volume element dV(z, H0, q0)/dz . It assumes H0 = 75 km/sec/Mpc, q0 = ½ and neglect evolution, an assumption likely to underestimate the counts for z>0.5. The numbers broadly agree with those in Wang (2000). The uncertainties in the observed rates do not warrant more sophisticated modeling. The uncertainties in these estimates are obvious and will not be discussed further.

## 4. TECHNOLOGICAL READINESS AND COSTS.

The LM technology is young so that it is legitimate to wonder whether it is sufficiently proven and robust to carry out the strip survey <u>now</u>. The answer is yes. LMs have been extensively tested in the Laval Liquid Mirror laboratory. Optical shop tests of a 2.5-m LM show that it is diffraction limited (Borra, Content & Girard 1993, Girard & Borra 1997). Extensive tests of a 3.7-m diameter LM show that it is robust and reasonably insensitive to outside perturbations (Tremblay & Borra 2000). A 3-m LMT (the NASA Orbital Debris Observatory; http://www.sunspot.noao.edu/Nodo/nodo.html) has been built and operated by NASA near Cloudcroft, New Mexico, for several years to search for space debris. The instrument has performed flawlessly. Although the telescope was designed for space debris observations, was not optimized for astronomical observations and has a relatively inefficient thick CCD, it has been used to obtain over 100 nights of astronomical data (Hickson & Mulrooney 1997). Cabanac and Borra (1998) have published the results of astronomical research carried out with a subset of 29 nights obtained with the NODO telescope. Figure 4 in Cabanac and Borra (1998) shows an eye-catching sample of the data obtained. Their figure 2 summarizes the performance of the instrument, showing that the LMT was sufficiently robust to operate night after night in an astronomical environment.

   Costs and time-to-build are also important issues. Tremblay (1999) has documented in detail the costs of the materials and the man-hours needed to build a 3.7-m mirror, so that its cost (essentially two orders of magnitudes less than a glass mirror) can be reliably predicted. As a matter of fact a 4-m LM can be built in a few months at very low cost in a University environment (as the Laval 3.7-m was), for its construction mostly involves technically unskilled labor . More information can be found at: http://wood.phy.ulaval.ca/home.html. In particular, the construction of the Laval 3.7-m mirror is shown in considerable detail in a section of that web site

(http://wood.phy.ulaval.ca/construction/LMTPIC_0.html). The low construction and operating costs of an LMT are also well documented by the NODO experiment. A 4-m LM can be built in a few months, while it takes a few years from the time one orders a glass mirror to the moment that it is delivered.

## 5. DISCUSSION

Keeping the uncertain SN rates in mind, it is fair to expect that a 4-m survey will discover and obtain light curves for several thousands of Type Ia Supernovae per year to z=1. Given the low operating costs of a LMT observatory, it could continue for several years, gathering the large samples needed to obtain sufficient statistics and to fully understand the material at hand. Needless to say, the survey will also give an unprecedented sample of variable stars (including other types of supernovae and extragalactic objects). Coadding the nightly data will allow one to obtain a very deep survey (B,R,I around $26^{th}$ magnitude in 1 year) in the strip of sky. In 4 years, the survey would go beyond $27^{th}$ magnitude, as deep as the FORS deep field survey (Heidt et al. 2001), but in a far larger field. This is a major advantage over proposals that want to secure large batches of time on existing telescopes, for they would never get that many nights. The zenith observations would also have greater interest, than pencil surveys, for other fields of astronomy, for the strip samples a fairer sample of the sky. For example, the strip continuously samples our galaxy, from shining plane to shining plane through dark poles. The science that can be done with the deep coadded data is obvious to a professional astronomer and will not be discussed further.

Obviously one can go fainter and to larger redshifts by building larger mirrors and/or an array of LMTs. I have conservatively restricted the discussion to a 4-m, because it is a small extrapolation from the LM laboratory 3.7-m and the NODO 3-m; but the tests done on the 3.7-m (Tremblay & Borra 2000) indicate that it should be possible to build LMs having significantly larger diameters. An array of 5 LMTs each observing in a different band would give simultaneous UBVRI multicolor curves, which should minimize, if not totally eliminate, the need for spectroscopic confirmation of the supernova type. Redshifts of the host galaxies can be efficiently obtained from multiobject spectroscopy since the width of the strip (26 arcminutes) is well matched to the field of a typical multiobject spectrograph. Should spectroscopy of the SNe be necessary, the use of an LMT to discover and follow the light curves will save precious time on oversubscribed conventional telescopes that could be better used for spectroscopic follow-up.

The discussion has been conservatively restricted to what can be done <u>now</u>. However, the technology is young so that there is considerable room for improvement. For example, laboratory work shows that LMs can be tilted (Borra, Ritcey & Artigau 1999). This will increase enormously their usefulness. Since that article, unpublished laboratory work has confirmed and strengthened the ideas put forward in that article..

## 6. CONCLUSION

Our consideration of a 4-m LMT strip survey predicts discovery rates and z-distributions that compare favorably with those of other proposed surveys, like the pencil beam survey proposed by Wang (2000) and the SNAP survey. However, it will do that at a small fraction of the cost of those projects. Observations in a long strip, as opposed to pencil beams, have the further advantage of sampling the Universe to a much larger scale.

Building a 4-m LMT is a low-risk undertaking because the mirror is a small extrapolation from the 3.7-m mirror tested in the laboratory and the 3-m NODO LMT.

Unlike some of the other proposed surveys, LMT surveys can start very soon, for a 4-m LM can be built in a few months, unlike a glass mirror that takes years to anneal and polish. As a matter of fact, in principle, a SN survey almost as deep as the one considered here could start a few weeks from reading this article, if one equips the NODO 3-m with a borrowed high-performance thin CCD. The NODO space debris survey uses twilight time, leaving the remainder of the night available for other observations. The observations of Hickson & Mulrooney (1998) were obtained that way. The SN survey is also one of the key projects of the ILMT consortium, a 4-m LMT planned in the southern hemisphere (http://vela.astro.ulg.ac.be/themes/telins/lmt/index_e.html). At the time of this writing, the project has been substantially supported but still needs some additional funding. It could complement a northern hemisphere survey with NODO.

## ACKNOWLEDGMENTS

This work was supported by a grant from the Natural Sciences and Engineering Research Council of Canada.

Table 1

Broadband limiting magnitudes. The integration time of 120 seconds is given by the time it takes an object to cross the face of the detector

(4-m LMT with a 4KxK CCD mosaic having 0.4-arcsecond pixels, S/N = 5)

===============================================================

| 1 night (120 sec) | | | 6 nights (720 sec) * | | |
|---|---|---|---|---|---|
| B | R | I | B | R | I |
| 24.4 | 24.1 | 23.7 | 25.4 | 25.1 | 24.7 |

* The data from 6 consecutive nights are coadded

Table 2

Apparent R and I magnitude of a type Ia supernova at maximum light as function of redshift

==================================================================

| z | $m_R$ | $m_I$ |
|---|---|---|
| 1.0 | 25 | 23.8 |
| 0.8 | 23.8 | 23.0 |
| 0.6 | 22.7 | 22.5 |
| 0.4 | 21.8 | 21.6 |
| 0.2 | 20.2 | 20.2 |

Table 3

Predicted SNe Ia discovery rates in $\Delta z = 0.2$ bins.

===============================================================

| z | events/year |
|---|---|
| 0.2 | 500 |
| 0.4 | 1000 |
| 0.6 | 1500 |
| 0.8 | 1900 |
| 1.0 | 2000 |